\documentclass{ws-procs9x6}

\begin{document}
\newcommand{\TeV}{\,{\rm TeV}}
\newcommand{\GeV}{\,{\rm GeV}}
\newcommand{\MeV}{\,{\rm MeV}}
\newcommand{\keV}{\,{\rm keV}}
\newcommand{\eV}{\,{\rm eV}}
\newcommand{\Tr}{{\rm Tr}\!}
\renewcommand{\arraystretch}{1.2}
\newcommand{\be}{\begin{equation}}
\newcommand{\ee}{\end{equation}}
\newcommand{\bea}{\begin{eqnarray}}
\newcommand{\eea}{\end{eqnarray}}
\newcommand{\ba}{\begin{array}}
\newcommand{\ea}{\end{array}}
\newcommand{\bc}{\begin{center}}
\newcommand{\ec}{\end{center}}
\newcommand{\bmat}{\left(\ba}
\newcommand{\emat}{\ea\right)}
\newcommand{\bds}{\begin{description}}
\newcommand{\eds}{\end{description}}
\newcommand{\refs}[1]{(\ref{#1})}
\newcommand{\ler}{\stackrel{\scriptstyle <}{\scriptstyle\sim}}
\newcommand{\ger}{\stackrel{\scriptstyle >}{\scriptstyle\sim}}
\newcommand{\lag}{\langle}
\newcommand{\rag}{\rangle}
\newcommand{\ns}{\normalsize}
\newcommand{\cm}{{\cal M}}
\newcommand{\gr}{m_{3/2}}
\newcommand{\p}{\partial}
\newcommand{\bsg}{$b\rightarrow s + \g$}
\newcommand{\Bsg}{$\bar{B}\rightarrow X_s + \g$}
\newcommand{\atal}{{\it et al.}}
\newcommand{\cq}{\mathcal{Q}}
\newcommand{\cqt}{\widetilde{\mathcal{Q}}}
\newcommand{\wtlc}{\widetilde{C}}
\def\321{$SU(3)\times SU(2)\times U(1)$}
\def\tl{{\tilde{l}}}
\def\tL{{\tilde{L}}}
\def\bd{{\overline{d}}}
\def\tL{{\tilde{L}}}
\def\a{\alpha}
\def\b{\beta}
\def\bsg{$ b \rightarrow s + \g$}
\def\g{\gamma}
\def\c{\chi}
\def\d{\delta}
\def\D{\Delta}
\def\db{{\overline{\delta}}}
\def\Db{{\overline{\Delta}}}
\def\e{\epsilon}
\def\f{\frac}
\def\tn{-\frac{2}{9}}
\def\tt{\frac{2}{3}}
\def\l{\lambda}
\def\n{\nu}
\def\m{\mu}
\def\nt{{\tilde{\nu}}}
\def\p{\phi}
\def\P{\Phi}
\def\k{\kappa}
\def\x{\xi}
\def\r{\rho}
\def\s{\sigma}
\def\t{\tau}
\def\th{\theta}
\def\ne{\nu_e}
\def\nm{\nu_{\mu}}
\def\snui{\tilde{\nu_i}}
\def\la{{\makebox{\tiny{\bf loop}}}}
\def\ti{\tilde}
\def\ssc{\scriptscriptstyle}
\def\wtl{\widetilde}
\def\mp{\marginpar}
\def\und{\underline}
\renewcommand{\Huge}{\Large}
\renewcommand{\LARGE}{\Large}
\renewcommand{\Large}{\large}

\title{\Bsg\ in Supersymmetry without $R$-parity}


\author{R.~D. VAIDYA\footnote{\uppercase{S}peaker at the conference.}~ and~ O.C.W. Kong}
\address{Department of Physics, \\
National Central University, \\ 
Chung-Li 32054 TAIWAN} 

\maketitle

\abstracts{
We present a systematic analysis of the decay \Bsg\ at the 
leading log within the 
framework of Supersymmetry without $R$-parity. We point out some new contributions in
the form of bilinear-trilinear combination of $R$-parity violating (RPV) couplings 
that are
enhanced by large $\tan \beta $. We also improve by a few orders of magnitude, bounds on 
several combinations of RPV parameters.}

\section{Introduction}
The large number of talks on supersymmetry (SUSY) in this conference provides 
ample proof of the
inadequacy of standard model (SM) as a complete theory, and the appeal of 
SUSY as a most 
popular candidate for the physics beyond SM. In our opinion, the minimal supersymmetry standard model
with conserved $R$-parity, lacks the much needed solution to neutrino mass problem which is naturally
addressed in models with $R$-parity violation (RPV). However, the large number of 
{\it a priori} arbitrary RPV couplings must be constrained from phenomenology in all possible ways.
In this talk we shall discuss the influence of RPV on the decay channel
\Bsg\ . Being loop mediated rare decay, it is sensitive to physics beyond SM. It has already been
well measured by CLEO, BELLE, ALEPH and BABAR and hence can be used to put upper bounds on RPV
couplings. The experimental world average\cite{world_average} is
$Br\left[\bar{B} \rightarrow X_s + \g \;(E_{\g} > 1.6 GeV) \right]_{\ssc \mathrm{SM}} 
= (3.57 \pm 0.30) \times 10^{-4}$. Within $1 \sigma$ this matches very well with the
SM prediction\cite{bur-mis-02} of
$ (3.57 \pm 0.30) \times 10^{-4}$.
The good agreement between SM prediction and the
experimental number at $1 \s$ can be used to constrain the large number of {\it a priori}
arbitrary parameters of SUSY without $R$-parity.

There have been few studies on the process within the general framework of 
$R$-parity violation\cite{carlos,besmer}. 
Ref.\cite{carlos},  fails to consider the
additional 18 four-quark operators which, in fact, give the dominant contribution
in most of the cases. 
Ref.\cite{besmer} has considered a complete operator basis. However, we find their
formula for Wilson coefficient (WC) incomplete, and they do not report on the possibility
of a few orders of magnitude improvement on the bounds for certain combinations of
RPV couplings, as we present here\cite{tri-tri}. In fact, the particular type of 
contributions --- namely, the one from a combination of
a bilinear and a trilinear $R$-parity violating (RPV) parameters, we focused 
on \cite{bi-tri}, has not been studied in any detail before. Here we shall briefly report
the results. For the analytical details we refer the readers to\cite{tri-tri,bi-tri}.

We adopt an optimal phenomenological parametrization of  the full model 
Lagrangian -- the single single-{\it vev} parametrization.
It is essentially about choosing a basis
for Higgs and lepton superfields in which all the ``sneutrino" {\it vev} vanish. The formalism
is discussed at length in\cite{otto-gssm}.
\section{Formalism} The partonic transition \bsg\ is described by the magnetic penguin diagram.
Under the effective Hamiltonian approach, the corresponding WC 
of the standard $\mathcal{Q}_7$ operator has many RPV contributions at
the scale $M_{\!\ssc W}$. For example, we separate the contributions from
different type of diagrams as 
$C_7 = C^{\ssc W}_7 \,+ \,C^{\ssc \tilde g}_7 \, + \,C^{\ssc \chi^-}_7 \, 
+\, C^{\ssc \chi^{\mbox{\tiny 0}}}_7 \, +\, C^{\ssc \phi^-}_7 \,
+\, C^{\phi^{\mbox{\tiny 0}}}_7$ corresponding to W-boson, gluino, chargino, neutralino,
colorless charged-scalar and colorless neutral-scalar loops (for details please see\cite{tri-tri}). 
Apart from the 8 SM operators with additional contributions, 
we actually   have  to consider many more operators with admissible nonzero  WC 
at  $M_{\!\ssc W}$ resulting from the RPV couplings. These are the
chirality-flip counterparts $\cqt_7$ and $\cqt_8$ 
of the standard (chromo)magnetic 
penguins $\cq_7$ and $\cq_8$, and a whole list of 18 new relevant four-quark 
operators. For the lack of space, we list 8 important operators below.
\bea
\cqt_{9-11}
\!\! & =& \!\!
\left({\bar s}_{L\a}\,\g^{\m} \, b_{L\b}\right)
\, \left({\bar q}_{R\b}\,\g_{\m}\,q_{R\a}\right)\,,~q = d,s,b;  \nonumber\\
\cqt_{9-13} 
\!\! & =& \!\!
 \left({\bar s}_{R\a}\,\g^{\m}\,b_{R\b}\right)
\, \left({\bar q}_{L\b}\, \g_{\m}\,q_{L\a}\right)\,, ~q = d,s,b,u,c;  
\eea
and six more operators from $\l''$ couplings\cite{tri-tri}. 
The interplay among the full set of 28 operators is what 
makes the analysis complicated. The effect of the QCD corrections proved to be 
very significant even for the RPV parts.

After the QCD running  of WC from 
scale $M_W$ to $m_b$, dictated by $28 \times 28$  anomalous dimension matrix, 
the effective WC
are given as (at leading log order)\cite{tri-tri}  :
\begin{eqnarray}
\label{c_mb}
C_7^{\mathrm{eff}}(m_b)&=& 
-0.351 \; C_{2}^{\mathrm{eff}}(M_{\!\ssc W})
+0.665 \;  C_{7}^{\mathrm{eff}}(M_{\!\ssc W})
+0.093 \; C_{8}^{\mathrm{eff}} (M_{\!\ssc W})\nonumber\\
&&-0.198 \; C_{9}^{\mathrm{eff}}(M_{\!\ssc W}) 
-0.198 \;  C_{10}^{\mathrm{eff}} (M_{\!\ssc W})
-0.178 \;  C_{11}^{\mathrm{eff}}(M_{\!\ssc W})\;, \nonumber \\
[0.5cm]
{\wtl C}_7^{\mathrm{eff}}(m_b)&=&
0.381 \;  {\wtl C}_{1}^{\mathrm{eff}}(M_{\!\ssc W})
+0.665  \; {\wtl C}_{7}^{\mathrm{eff}}(M_{\!\ssc W})
+0.093  \; {\wtl C}_{8}^{\mathrm{eff}}(M_{\!\ssc W})\nonumber\\
&&-0.198  \; {\wtl C}_{9}^{\mathrm{eff}}(M_{\!\ssc W}) 
-0.198 \;  {\wtl C}_{10}^{\mathrm{eff}}(M_{\!\ssc W})
-0.178  \; {\wtl C}_{11}^{\mathrm{eff}}(M_{\!\ssc W})\nonumber\\
&&+0.510 \;  {\wtl C}_{12}^{\mathrm{eff}}(M_{\!\ssc W})
+0.510 \; {\wtl C}_{13}^{\mathrm{eff}}(M_{\!\ssc W})
+0.381 \; {\wtl C}_{14}^{\mathrm{eff}}(M_{\!\ssc W})\nonumber\\
&&-0.213 \; {\wtl C}_{16}^{\mathrm{eff}}(M_{\!\ssc W})\;.
\end{eqnarray}
The branching fraction for $Br (b \rightarrow s + \g )$  is expressed through the 
semi-leptonic decay $b \rightarrow u|c e{\bar \nu}$ 
(so that the large
bottom mass dependence $( \sim m^5_b)$ and uncertainties in CKM elements
cancel out)
with $Br_{\mathrm{exp}} (b \rightarrow u|c\,e\,{\bar \nu_e}) = 10.5 \%$
and $\Gamma (b \rightarrow s \g) \propto (\,|C^{\mathrm{eff}}_7 (\m_b)|^2 + 
|\widetilde{C}^{\mathrm{eff}}_7(\m_b)|^2 )$.
Note that we have also to include RPV contributions to the semi-leptonic rate
for consistency.
\section{Results: Impact of bilinear-trilinear combination of parameters}
{\it Analytical Appraisal.--}We implement our (1-loop) calculations using mass eigenstate 
expressions, hence free from the commonly adopted mass-insertion 
approximation. While a trilinear RPV parameter gives a vertex, a bilinear 
parameter now contributes only through mass mixing matrix elements characterizing 
the effective couplings of the mass eigenstate running inside the loop. The $\mu_i$'s
are involved in fermion, as well as scalar mixings. There are also the
corresponding soft bilinear  $B_i$ parameters involved only in scalar 
mixings\cite{otto-gssm}. Combinations of $\mu_i$'s and  $B_i$'s with the trilinear
$\lambda^{\!\prime}_{ijk}$ parameters are our major focus.

There are two kinds of $B_i$-$\l'$ combinations that contribute to \bsg\ at 1-loop: 
(a) $B^*_i \l^{'}_{ij2}$, and (b) $B_i \l^{'*}_{ij3}$. These involve quark-slepton loop diagrams.
Case (a) leads to the $b_{\!\ssc L} \rightarrow s_{\!\ssc R}$ transition (where SM and MSSM 
contribution is extremely suppressed) whereas case (b) leads to SM-like $b_{\!\ssc R} 
\rightarrow s_{\!\ssc L}$ transition. For the purpose of illustration, we will assume a 
degenerate slepton spectrum and take the sleptonic index $i=3$ as a representative. 
For the $j$ values,
the charged loop contributions are still possible by invoking CKM mixings. 
Consider the contribution of case (a) with $|B_3^* \l'_{332}|$ to the 
${\wtl C}_7$, for instance. Through the extraction of the bilinear mass mixing
effect under a perturbative diagonalization of the mass matrices\cite{otto-gssm}, 
we obtain\cite{bi-tri},
\bea
{\wtl C}^{\ssc \phi^-}_7  & \approx &
 \frac{ - |V^{tb}_{\!\mbox{\tiny CKM}}|^2 |B_3^* \l'_{332}| }
{M^2_s}
\left\{
y_b   \tan_{\b}
 \left[ 	F_{12}   (x_t)	
\right] 
+ \! \f{y_t  m_t}{m_b} 
\left[ 	F_{34} (x_t)
\right] 		\right\}\nonumber\\
{\wtl C}^{\phi^{\mbox{\tiny 0}}}_7  &\approx &
\f{-2q_d \, y_b \, |B_3^*  \l'_{332}| \tan_{\b}}{M^2_s M_{\!\scriptscriptstyle S}^2}
F_1 (x_b) 
\eea
for the  charged and neutral colorless scalar loop, respectively. Here $x_t$ stands for
$({m}_{\!\scriptscriptstyle t}^2 / M_{\!\scriptscriptstyle \tilde{\ell}}^2)
$ with an obvious replacement for $x_b$. Here $F_{12(34)}(x_t) = q_u F_{1(3)}(x_t) + F_{2(4)}(x_t)$
where $F_i$ are the well known loop functions\cite{tri-tri}.
In the above equations, proportionality to $\tan\!{\b}$ shows the 
importance of these contributions in the large $\tan\!{\b}$ limit. The $M_s^2$,
$M_{\!\scriptscriptstyle \tilde{\ell}}^2$, $M_{\!\scriptscriptstyle S}^2$,
are all scalar (slepton/Higgs) mass parameters.
The term proportional to $y_t$ above has chirality flip inside the loop. Thinking in terms of
the electroweak states, it is easy to appreciate that the loop diagram giving a corresponding
term for ${\wtl C}^{\phi^{\mbox{\tiny 0}}}_7$ ({\it cf.} involving
$\widetilde{\mathcal{ N}}_{\!\scriptscriptstyle nm3}^{\!\scriptscriptstyle L}\,
 \widetilde{\mathcal{ N}}_{\!\scriptscriptstyle nm2}^{\!\scriptscriptstyle R^*}$)
requires a Majorana-like scalar mass insertion, which has to arrive from other RPV 
couplings\cite{otto-gssm}. 
In the limit of perfect mass degeneracy between the scalar and 
pseudoscalar part (with no mixing) of multiplet, it vanishes. Dropping this smaller
contribution, together  with the difference among the Inami-Lim loop functions and the
fact that the charged loop has more places to attach the photon (with also larger charge
values) adding up, we expect the ${\wtl C}^{\ssc \phi^-}_7$ to be larger than
${\wtl C}^{\phi^{\mbox{\tiny 0}}}_7$. We corroborate these features in our numerical 
study.

{\it Numerical Results.--} We take non-vanishing values
for relevant combinations of a bilinear and a trilinear RPV parameters one at a time,
and stick to real values only. Our model choice for parameters is 
(with all mass dimensions
given in GeV):  squark masses 300, down-type Higgs mass 300, $\m_{\ssc 0} = -300 $ 
sleptons mass 150 and gaugino mass $M_2 = 200$ (with $M_1 = 0.5 M_2$ and 
$M_3 = 3.5M_2$), $\tan\!\b = 37$ and $A$ parameter 300. 
We impose the experimental number to obtain bounds
for each combination of RPV parameters independently (given in Table I). 
Consider, for instance, the case (b)
combination $|B_{\ssc 3}\l^{'*}_{\ssc 323}|$.  We obtain a bound of $5.0 \cdot 10^{-5}$, when
normalized by a factor of $\m^2_{\ssc 0}$. Since this is a $b_{\!\ssc R} \rightarrow s_{\!\ssc L}$
transition, the RPV contribution interferes with the SM as well as the MSSM contribution.
Over and above
the loop contributions there are contributions coming from four-quark 
operator with $C_{11}$ ($\propto y_b$) which is stronger than 
the other two four-quark quark coefficients ${\wtl C}_{10,13} \propto y_s$.
Since the neutral scalar loop contribution is proportional
to the loop function $F_1$ (which is of order .01), it is suppressed compared to 
current-current contributions. Also here the charged scalar contribution comes only 
with chirality flip inside the loop and has a CKM suppression. So the current-current is 
dominant. It has a more subtle role to play when one writes the regularization
scheme-independent $C^{\mathrm{eff}}_7 = C_7 -C_{11}$ at scale $M_{\ssc W}$(see\cite{tri-tri}).
Due to dominant and negative
sign chargino contribution (because $A_t\, \m_{\ssc 0} < 0$), 
the positive sign $C_{11}$ interferes
constructively with $C_7$ and enhances the rate.These features can be verified from
Fig.1 of Ref.\cite{bi-tri}. We have done the similar analytical and numerical exercise
for all possible combinations of bilinear and trilinear couplings and quote the
relevant bounds obtained for the first time in Table\ref{bounds}.

{\it Conclusions. ---} We have systematically studied the influence of the 
combination of bilinear-trilinear RPV parameters on the decay \bsg\ analytically as 
well as numerically. 
These contributions are enhanced 
by large $\tan\!\b$. We also demonstrate the importance of QCD corrections and
 obtain strong bounds on several combinations of RPV parameters for the first time. 
Numerical study
has also been performed on combinations of trilinear parameters\cite{tri-tri}. 
We quote here a few
exciting bounds under a similar sparticle spectrum. 
For instance $|\l'_{i33}\cdot\l^{'*}_{i23}|$ for $i=2,3$
should be less than $1.6\cdot10^{-3}$ to be compared with rescaled existing bound of
$2\cdot 10^{-2}$. 
\begin{table}[t]
\tbl{Our Bounds for the absolute value of products of RPV couplings.}
{\small
\vspace{0.3cm}
\label{bounds}
\begin{tabular}{llllll}\hline
{\bf Product } & {\bf Bound} & {\bf Product} & {\bf bound} & {\bf Product} & {\bf bound}
 \\\hline  
$\f{B_i \cdot \l'_{i23}}{\m^2_{\ssc 0}}$& $5.0\cdot 10^{-5}$&
$\f{B_i \cdot \l'_{i12}}{\m^2_{\ssc 0}}$& $4.5\cdot 10^{-2}$&
$\f{\m_i \cdot \l'_{i23}}{\m_{\ssc 0}}$& $2.2\cdot 10^{-3}$\\
$\f{B_i \cdot \l'_{i32}}{\m^2_{\ssc 0}}$& $7.4\cdot 10^{-3}$&
$\f{B_i \cdot \l'_{i22}}{\m^2_{\ssc 0}}$& $6.5\cdot 10^{-2}$&
$\f{\m_i \cdot \l'_{i32}}{\m_{\ssc 0}}$& $1.0\cdot 10^{-2}$\\
$\f{B_i \cdot \l'_{i33}}{\m^2_{\ssc 0}}$& $2.3\cdot 10^{-3}$&
$\f{B_i \cdot \l'_{i13}}{\m^2_{\ssc 0}}$& $8.0\cdot 10^{-2}$&
$\f{\m_i \cdot \l'_{i33}}{\m_{\ssc 0}}$& $8.0\cdot 10^{-2}$\\\hline
\end{tabular}}
\end{table} 
\section*{Acknowledgment}
We would like to thank 
National Science Council of Taiwan and Physics department (NCU) for the 
support under post-doc grant number 
NSC 92-2811-M-008-012. 
\bibliographystyle{plain}

\end{document}